\documentclass[twocolumn,showpacs,superscriptaddress,amsmath,amssymb,aps,pra]{revtex4-1}
\usepackage{graphicx}
\usepackage{CJKutf8}
\usepackage{dcolumn}
\usepackage{amsmath,bm}
\usepackage{epstopdf}
\usepackage[mathlines]{lineno}
\usepackage{color}
\usepackage[colorlinks=true,linkcolor=blue,citecolor=magenta,urlcolor=cyan]{hyperref}

\begin{document}
\title{Energy spectrum, the spin polarization, and the optical selection rules of the Kronig-Penney superlattice model with spin-orbit coupling}
\author{Rui\! Li~(\begin{CJK}{UTF8}{gbsn}李睿\end{CJK})}
\email{ruili@ysu.edu.cn}
\affiliation{Key Laboratory for Microstructural Material Physics of Hebei Province, School of Science, Yanshan University, Qinhuangdao 066004, China}
\affiliation{Quantum Physics and Quantum Information Division, Beijing Computational Science Research Center, Beijing 100193, China}

\begin{abstract}
The Kronig-Penney model, an exactly solvable one-dimensional model of crystal in solid physics, shows how the allowed and forbidden bands are formed in solids. In this paper, we study this model in the presence of both strong spin-orbit coupling and the Zeeman field. We analytically obtain four transcendental equations that represent an implicit relation between the energy and the Bloch wavevector. Solving these four transcendental equations, we obtain the spin-orbital bands exactly. In addition to the usual band gap opened at the boundary of the Brillouin zone, a much larger spin-orbital band gap is also opened at some special sites inside the Brillouin zone. The $x$-component of the spin-polarization vector is an even function of the Bloch wavevector, while the $z$-component of the spin-polarization vector is an odd function of the Bloch wavevector. At the band edges, the optical transition rates between adjacent bands are nonzero.
\end{abstract}
\date{\today}
\maketitle

\section{introduction}

Semiconductor materials with strong spin-orbit coupling (SOC) have attracted considerable interest in recent years because of their potential applications in both condensed-matter physics and quantum-information processing~\cite{Hasan,Fu,Liu}. For example, a topological insulator phase has been discovered in strong spin-orbit coupled quantum-well structures~\cite{Konig}, and a strong spin-orbit coupled semiconductor nanowire in proximity to a {\it s}-wave superconductor can realize an 1D topological superconductor~\cite{Lutchyn,Oreg}. Also, a spin qubit confined in a semiconductor quantum dot with strong SOC has the advantage of being electrically manipulable~\cite{Nowack,Nadj1,Rashba1,Golovach,Li,Echeverria,Nowak,Tokura,Hung}.

The emergence of SOC in semiconductor materials is because of the lacking of space-inversion symmetry. Bulk inversion asymmetry leads to Dresselhaus SOC~\cite{Dresselhaus}, and structure inversion asymmetry leads to Rashba SOC~\cite{Bychkov}. Moreover, Rashba SOC can be tuned to some extent by an external electric field~\cite{Nitta}, such that it is possible to tune a quantum system to the strong SOC regime. In the presence of SOC, the spin degree of freedom is mixed with the orbital degree of freedom of the electron, such that it is usually difficult to clarify the strong SOC effects.

The Kronig-Penney (KP) model~\cite{Kronig} is an 1D model of a crystal that shows how the electrons are dispersed into allowed and forbidden bands~\cite{Kittel}. The KP model is important in condensed-matter physics due to its exact solvability. It is interesting to ask, when nontrivial SOC is presented in the KP model, whether the resulting spin-orbital superlattice model is still exactly solvable and what are the consequences. The energy spectrum~\cite{Khomitsky,Foldi,Vurgaftman}, the optical properties~\cite{Khomitsky2,Perov}, and the transport properties~\cite{Wang,Kleinert,Foldi2} of various spin-orbital superlattice models have received considerable attention in recent years.

In this paper, we study the KP superlattice model with SOC, where an electron moves in an 1D periodic $\delta$ potential~\cite{Marinescu} in the presence of both SOC and the Zeeman field. We derive analytically four transcendental equations, which represent an implicit relation between the energy and the Bloch wavevector. By solving these transcendental equations, the spin-orbital bands~\cite{ZhangII,Thorgilsson} are obtained exactly. In addition to the band gap usually opened at the boundary of the Brillouin zone~\cite{Kronig}, a much larger spin-orbital band gap can also be opened at some special sites inside the Brillouin zone. The spin-orbital band gap is a representative character of the spin-orbital superlattice system. The norm of the spin polarization vector has a large jump at the boundary of the Brillouin zone, while it only has a small jump at the sites where the spin-orbital gap opened. Finally, we discuss the optical selection rules between adjacent bands. There is a finite optical transition rate between adjacent bands at the band edge.

\section{The model and the boundary conditions}
The model we are interested in is the KP superlattice model~\cite{Kittel} in the presence of both the SOC and an external Zeeman field. The Hamiltonian under consideration reads (in all of the following, we set $\hbar=1$)
\begin{equation}
H=-\frac{1}{2m}\partial^{2}_{x}-i\alpha\sigma^{z}\partial_{x}+\Delta\sigma^{x}+V_{0}a\sum^{N-1}_{n=0}\delta(x-na),\label{eq_Hamiltonain}
\end{equation}
where $m$ is the effective electron mass, $\alpha$ is the Rashba SOC strength, $\Delta=g_{e}\mu_{B}B/2$ is half of the Zeeman splitting (with $g_{e}$, $\mu_{B}$, and $B$ being the effective g-factor, the Bohr magneton, and the Zeeman field, respectively), and the last term is the periodic $\delta$ potential with lattice periodicity $a$ and potential height $V_{0}$.

We now analyze the boundary conditions of our model. Due to the lattice periodicity, we only need to consider the boundary conditions at the site $x=0$ because the boundary conditions at other sites $x=a$, $2a$,$\ldots$ are just the same as that at the site $x=0$. First, due to the continuous property of the wave function, we have one boundary condition:
\begin{equation}
\Psi(+0)=\Psi(-0),\label{eq_boundary_I}
\end{equation}
where $\Psi(x)$ is the eigenfunction of Hamiltonian (\ref{eq_Hamiltonain}). Second, integrating the Schr\"odinger equation in the vicinity of the site $x=0$: $\underset{\varepsilon\rightarrow\,0}{\rm lim}\int^{\varepsilon}_{-\varepsilon}dx(H-E)\Psi=0$, we have the other boundary condition:
\begin{equation}
\Psi'(+0)-\Psi'(-0)=2mV_{0}a\Psi(0),\label{eq_boundary_II}
\end{equation}
where $\Psi'(\pm0)$ is the first derivative of the eigenfunction at the site $x=\pm0$.

As one can find out, the boundary conditions of this model look the same as that of the bare KP model~\cite{Kronig} (without a spin degree of freedom). However, the wavefunction here, $\Psi(x)=[\Psi_{1}(x),\Psi_{2}(x)]^{\rm T}$, has two components, such that the boundary conditions (\ref{eq_boundary_I}) and (\ref{eq_boundary_II}) actually contain four equations.

It should be noted that in all of our following calculations, unless otherwise stated, we have chosen InSb nanowire~\cite{Nadj2, Li2} as our superlattice material, and the detailed parameters of our model are given in Tab.~\ref{Tab_Parameters}.

\begin{table}
\caption{\label{Tab_Parameters}The parameters of the InSb quantum wire superlattice used in our calculations}
\begin{ruledtabular}
\begin{tabular}{cccccc}
$g_{e}$~\cite{Nadj2}&B(T)&$m/m_{e}$\footnote{$m_{e}$ is the electron mass}~\cite{Nadj2}&$x_{\rm so}$\footnote{$x_{\rm so}=\hbar/(m\alpha)$ is the spin-orbit length} (nm)&$a$ (nm)&$V_{0}~({\rm meV})$\\
-50.6&0.4&0.0136&50,~200&100&0.5,~2
\end{tabular}
\end{ruledtabular}
\end{table}

\section{The spin-orbital bands}
\begin{figure}
\includegraphics{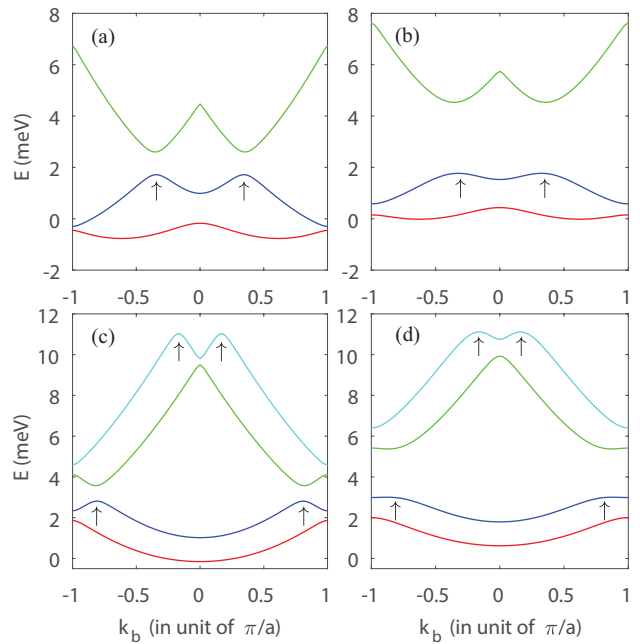}
\caption{\label{bandstructure}The band structure of the KP superlattice model with SOC. (a) The lowest three bands for $x_{\rm so}=50$ nm and $V_{0}=0.5$ meV. (b) The lowest three bands for $x_{\rm so}=50$ nm and $V_{0}=2$ meV. (c) The lowest four bands for $x_{\rm so}=200$ nm and $V_{0}=0.5$ meV. (d) The lowest four bands for $x_{\rm so}=200$ nm and  $V_{0}=2$ meV. The $k$ sites where the gap opened because of the SOC effects are marked with arrows.}
\end{figure}
We first solve the bulk spectrum and the corresponding bulk wave functions for our Hamiltonian (\ref{eq_Hamiltonain}) [for details see Appendix~\ref{appendix_A}]. Each obtained bulk wave function does not satisfy the desired boundary conditions (\ref{eq_boundary_I}) and (\ref{eq_boundary_II}). However, a linear combination of the four bulk wave functions can fulfill the boundary conditions~\cite{Bulgakov,Tsitsishvili}. Following this method,
in different energy regions, we have obtained four transcendental equations, i.e., Eqs.~(\ref{eq_transc1}), (\ref{eq_transc2}), (\ref{eq_transc3}), and (\ref{eq_transc4}) [for details see Appendix~\ref{appendix_B}], which represent an implicit relation between the energy and the Bloch wavevector. Solving these four transcendental equations, we can get the complete energy band for the KP superlattice model with SOC.

We do not make any approximation or assumption in deriving these transcendental equations, such that our results are general and exact. Here, we take the InSb nanowire~\cite{Nadj2, Li2} as an example to show the results we obtained. Figures~\ref{bandstructure}(a) and (b) show the band structure in the first Brillouin zone when our model is in the strong SOC regime $x_{\rm so}=50$ nm, i.e., $m\alpha^{2}>g_{e}\mu_{B}B$. Figures~\ref{bandstructure}(c) and (d) show the band structure in the first Brillouin zone when our model is in the weak SOC regime $x_{\rm so}=200$ nm, i.e., $m\alpha^{2}<g_{e}\mu_{B}B$. Meanwhile, Figs.~{\ref{bandstructure}}(a) and (c) give the band structure when the potential barrier is small, $V_{0}=0.5$ meV, and Figs.~{\ref{bandstructure}}(b) and (d) give the band structure when the potential barrier is relatively large, $V_{0}=2$ meV. At first glance, the band structure in the strong SOC regime is distinctly different from that in the weak SOC regime. Actually, this difference can be traced back to the difference originating from the bulk spectrum [see Fig.~\ref{fig_bulkspectrum}].

As expected, at the boundary sites of the Brillouin zone, e.g., $k_{b}=\pm\pi/a$, a band gap is opened because of the weak periodic potential $V_{0}$ term. This band gap is usually small, and it can be calculated using perturbation theory~\cite{Kittel}. In particular, a much larger band gap also opened at some special $k$ sites, which are marked as arrows in the figures [see Fig.~\ref{bandstructure}]. We call this gap the spin-orbital gap~\cite{ZhangII,Thorgilsson},  because its emergence is due to the interplay between the SOC and the periodic potential. As can be seen from Figs.~\ref{bandstructure}(b) and (d), the spin-orbital gap is much larger (several times larger) than the gap that opened at the boundary of the Brillouin zone. The spin-orbital gap is a representative character of the spin-orbital superlattice system. Obviously, the large spin-orbital gaps shown in the figures cannot be obtained using perturbation calculations, especially for the cases when the potential barrier is relatively large [see Figs.~{\ref{bandstructure}}(b) and (d)].

Band engineering is an interesting topic for the spin-orbital supperlattice system. Many parameters, e.g., the lattice constant $a$, the potential barrier $V_{0}$, and the spin-orbit strength $\alpha$, are externally tunable, such that it is possible to produce a nearly flat band~\cite{Zhang,WuLi} or well-separated spin-orbital bands [see Fig.~{\ref{bandstructure}}(b)]. It is also possible to engineer the topological superconductivity when the spin-orbital superlattice is in proximity to a {\it s}-wave superconductor~\cite{Lu,Levine}.

\section{The spin polarization}
\begin{figure}
\includegraphics{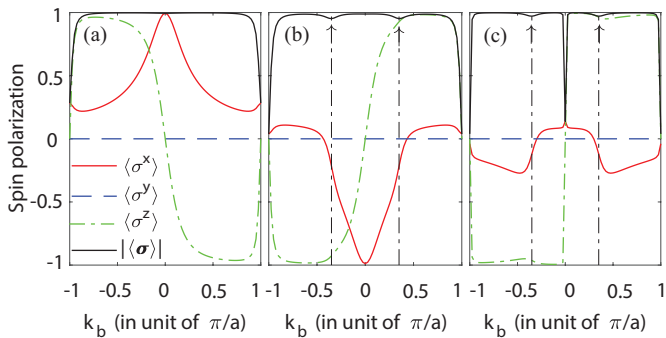}
\caption{\label{fig_polarstr}The spin expectation value in the first Brillouin zone calculated for the bands given in Fig.~{\ref{bandstructure}}(a). (a) The results for the first band. (b) The results for the second band. (c) The results for the third band. The arrows here mark the same $k$ sites as in Fig.~{\ref{bandstructure}}(a).}
\end{figure}
\begin{figure}
\includegraphics[width=8.5cm]{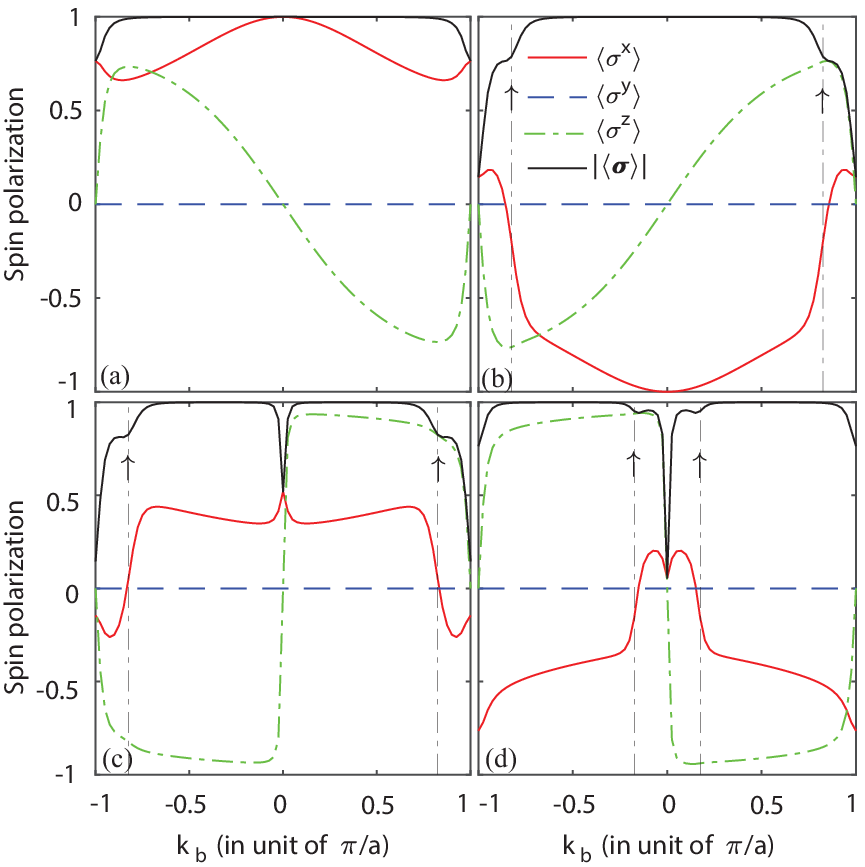}
\caption{\label{fig_polarweak}The spin expectation value in the first Brillouin zone calculated for the bands given in Fig.~{\ref{bandstructure}}(c). (a) The results for the first band. (b) The results for the second band. (c) The results for the third band. (d) The results for the forth band. The arrows here mark the same $k$ sites as in Fig.~{\ref{bandstructure}}(c).}
\end{figure}
In the presence of the SOC and the Zeeman field, the spin operator is no longer a good quantum number, i.e., neither a single operator ($\sigma^{z}$ or $\sigma^{x}$) nor a linear combination of $\sigma^{z}$ and $\sigma^{x}$ is a conserved quantity in Hamiltonian (\ref{eq_Hamiltonain}). The interplay of the SOC and the periodic potential produces complicated and well separated spin-orbital bands [see Fig.~\ref{bandstructure}].  Each $k_{b}$ site of a given band can only accommodate one electron. Although spin is not a conserved quantity in our superlattice system, it is still of interest to know the spin polarization direction when electrons occupy a given band. The spin polarization also reflects the magnetic properties of the superlattice system~\cite{Demikhovskii}. The spin polarization at a given $k_{b}$ site of a given band is described by the expectation value of the spin operator,
\begin{equation}
\langle\sigma^{x,y,z}\rangle=\int\,dx\Psi^{\dagger}_{n,k_{b}}(x)\sigma^{x,y,z}\Psi_{n,k_{b}}(x),
\end{equation}
where $\Psi^{\dagger}_{n,k_{b}}(x)$ is the Bloch function of a given band $n$. We calculate the spin polarization in the first Brillouin zone for the KP superlattice system in both the strong SOC regime, $x_{\rm so}=50$ nm [see Fig.~\ref{fig_polarstr}], and the weak SOC regime, $x_{\rm so}=200$ nm [see Fig.~\ref{fig_polarweak}]. It should be noted that because the spin operator is no longer a good quantum number, the norm of the spin polarization vector is less than $1$~\cite{Sherman} in our model. The norm of the spin polarization vector is equal to $1$ for systems in which spin is a good quantum number. First, at the boundary sites of the Brillouin zone where the traditional band gaps are opened, the norm of the spin polarization vector has a large jump, $|\langle\boldsymbol\sigma\rangle_{k_{b}=\pm\pi/a,\pm2\pi/a}|<1$ (see Figs.~\ref{fig_polarstr} and \ref{fig_polarweak}). Second, at some special $k_{b}$ sites where the spin-orbital gaps are opened (see the sites marked with arrows in Figs.~\ref{fig_polarstr} and \ref{fig_polarweak}), the norm of the spin polarization vector only has a minor jump, $|\langle\boldsymbol\sigma\rangle_{k_{b}={\rm marked~sites}}|\approx1$.

The components of the spin polarization vector have the following simple property
\begin{eqnarray}
\langle\sigma^{x}\rangle_{-k_{b}}&=&\langle\sigma^{x}\rangle_{k_{b}},\nonumber\\
\langle\sigma^{z}\rangle_{-k_{b}}&=&-\langle\sigma^{z}\rangle_{k_{b}}.\label{Eq_polarization}
\end{eqnarray}
The x-component of the spin polarization vector is an even function of $k_{b}$, and the z-component of the spin polarization vector is an odd function of $k_{b}$. Because of the periodic property in the Brillouin zone, $\langle\sigma^{z}\rangle_{k_{b}}=\langle\sigma^{z}\rangle_{k_{b}+2\pi/a}$, such that at the boundary sites $k_{b}=0,\pm\pi/a$, the spin polarization along the $z$ direction is exactly zero,
\begin{equation}
\langle\sigma^{z}\rangle_{k_{b}=0,\pm\pi/a}=0.
\end{equation}
There is no $\sigma^{y}$ term in our model, such that the y-component of the spin polarization vector is exact zero for all the $k_{b}$ sites $\langle\sigma^{y}\rangle_{k_{b}}=0$~\cite{Demikhovskii}. When the superlattice system contains $N$ electrons and the bands are occupied up to the Fermi energy $E_{f}$, it is interesting to show that there always exists a net $x$-polarization for the whole system $\sum_{E_{k_{b}}<E_{f}}\langle\sigma^{x}\rangle_{k_{b}}\neq0$, while there is no net $z$-polarization $\sum_{E_{k_{b}}<E_{f}}\langle\sigma^{z}\rangle_{k_{b}}=0$~\cite{Khomitsky}. This property can also be deduced from Eq.~(\ref{Eq_polarization}).

\section{The optical selection rules}
\begin{figure}
\includegraphics{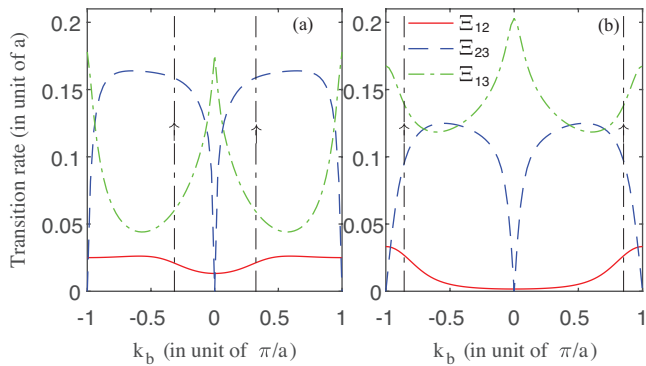}
\caption{\label{fig_selectionrule}The optical transition rate between different bands in the first Brillouin zone. (a) The results for the bands given in Fig.~\ref{bandstructure}(b). (b) The results for the bands given in Fig.~\ref{bandstructure}(d). The arrows here mark the same $k$ sites as in Figs.~{\ref{bandstructure}}(b) and (d).}
\end{figure}
An electron in a fully occupied valence band can absorb a photon and then be excited to the high-energy conduction band; the formation of the exciton state in a semiconductor is an illustration. To understand the optical properties~\cite{Perov} of the spin-orbital superlattice system, it is instructive to know the electric-dipole transition rate between different bands~\cite{Demikhovskii},
\begin{equation}
\Xi_{nm}(k_{b})=\left|\int\,dx\Psi^{\dagger}_{n,k_{b}}(x)x\Psi_{m,k_{b}}(x)\right|.
\end{equation}
where $n$ and $m$ are the band index marking the different bands. Figure~\ref{fig_selectionrule}(a) gives the optical transition rate between different bands when the superlattice is in the strong SOC regime. Figure~\ref{fig_selectionrule}(b) gives the optical transition rate between different bands when the superlattice is in the weak SOC regime. Generally, because of the mixing of the spin and the orbital degrees of freedom due to the SOC, except at the boundary of the Brillouin zone, e.g., $k_{b}=0$, $\pm\pi/a$, the transition rate between two arbitrary spin-orbital bands is nonzero. As can also be seen from the figure, $\Xi_{23}(k_{b}=0,\pm\pi/a)=0$ in both the strong and the weak SOC regimes. This is because the Bloch functions at these sites have an additional $Z_{2}$ symmetry~\cite{Debald} as pointed out in the literature~\cite{Demikhovskii}. The Bloch functions at the boundary of the Brillouin zone can be labeled with $\sigma^{x}\mathcal{P}=\pm1$, where $\mathcal{P}$ is the parity operator. The optical transition rate is zero between those Bloch functions with the same $Z_{2}$ symmetry. It should be noted that at the band edges, which are marked with arrows in Fig.~\ref{fig_selectionrule}, the optical transition rates are generally nonzero.

\section{Summary}
In summary, in this paper we have studied in detail the band structure, the spin polarization, and the optical selection rules of the KP superlattice model with SOC. We have analytically obtained four transcendental equations, i.e., Eqs.~(\ref{eq_transc1}), (\ref{eq_transc2}), (\ref{eq_transc3}), and (\ref{eq_transc4}), which describe an implicit expression between the energy $E$ and the Bloch wavevector $k_{b}$. In addition to the usual band gap opened at the boundary of the Brillouin zone, a larger spin-orbital band gap can also be opened at some special sites inside the Brillouin zone. With the exact energy spectrum and the corresponding Bloch functions obtained, we are able to calculate the spin polarization of a given band and the optical selection rules between different bands. The norm of the spin polarization has a jump at the $k_{b}$ sites where the gap opened. The jump is larger at the boundary of the Brillouin zone than that at the sites where the spin-orbital gaps opened. At the boundary of the Brillouin zone, some optical transitions are forbidden when the corresponding Bloch functions have the same $Z_{2}$ symmetry.

\section*{Acknowledgements}
We thank Zhaoxin Liang for useful discussion. This work is supported by National Natural Science Foundation of China Grant No.~11404020 and Postdoctoral Science Foundation of China Grant No.~2014M560039.

\appendix

\section{The bulk spectrum and the bulk wave functions}\label{appendix_A}
Due to the special property of the $\delta$ function, the periodic potential is zero everywhere except at the boundary sites. To find the energy spectrum of Hamiltonian (\ref{eq_Hamiltonain}), our first step is to find the bulk spectrum and the corresponding bulk wave functions, where the bulk Hamiltonian reads $H_{\rm b}=-\frac{1}{2m}\partial^{2}_{x}-i\alpha\sigma^{z}\partial_{x}+\Delta\sigma^{x}$~\cite{YunLi,Gambetta}. There exist two kinds of bulk wave functions, i.e., the plane-wave solution and the exponential function solution~\cite{Bulgakov,Tsitsishvili}. First, we consider the plane-wave solution. The bulk wave function can be assumed as
\begin{equation}
\Psi_{\rm b}(x)=e^{ikx}\left(\begin{array}{c}\chi_{1}\\\chi_{2}\end{array}\right),
\end{equation}
where $\chi_{1,2}$ are the coefficients to be determined. The bulk Schr\"odinger equation $(H_{\rm b}-E_{\rm b})\Psi_{\rm b}=0$ gives us the following matrix equation
\begin{equation}
\left(\begin{array}{cc}\frac{k^{2}}{2m}-E_{\rm b}+\alpha\,k&\Delta\\\Delta&\frac{k^{2}}{2m}-E_{\rm b}-\alpha\,k\end{array}\right)\cdot\left(\begin{array}{c}\chi_{1}\\\chi_{2}\end{array}\right)=0.\label{eq_bulkschrodinger1}
\end{equation}
Setting the determinant of the matrix (on the left side of the above equation) equal to zero, we get the bulk spectrum
\begin{equation}
E^{\pm}_{\rm b}=\frac{k^{2}}{2m}\pm\sqrt{\alpha^{2}k^{2}+\Delta^{2}}.\label{eq_bulkspectrum}
\end{equation}
Substituting the bulk energy $E^{\pm}_{\rm b}$ in Eq.~(\ref{eq_bulkschrodinger1}) with the above results, we obtain the corresponding bulk wave functions:
\begin{equation}
\Psi^{+}_{\rm b}=\left\{\begin{array}{c}e^{ikx}\left(\begin{array}{c}\cos\frac{\theta}{2}\\\sin\frac{\theta}{2}\end{array}\right)\\
e^{-ikx}\left(\begin{array}{c}\sin\frac{\theta}{2}\\\cos\frac{\theta}{2}\end{array}\right)\end{array}\right.,
\Psi^{-}_{\rm b}=\left\{\begin{array}{c}e^{ikx}\left(\begin{array}{c}\sin\frac{\theta}{2}\\-\cos\frac{\theta}{2}\end{array}\right)\\
e^{-ikx}\left(\begin{array}{c}\cos\frac{\theta}{2}\\-\sin\frac{\theta}{2}\end{array}\right)\end{array}\right.,\label{eq_branchwave1}
\end{equation}
where $\theta\equiv\theta(k)=\arctan\left[\Delta/(\alpha\,k)\right]$. Because $E^{\pm}_{\rm b}$ is an even function of $k$, there are two degenerate bulk wave functions, i.e., the left-moving $+k$ and the right-moving $-k$ wave functions.

\begin{figure}
\includegraphics[width=8.5cm]{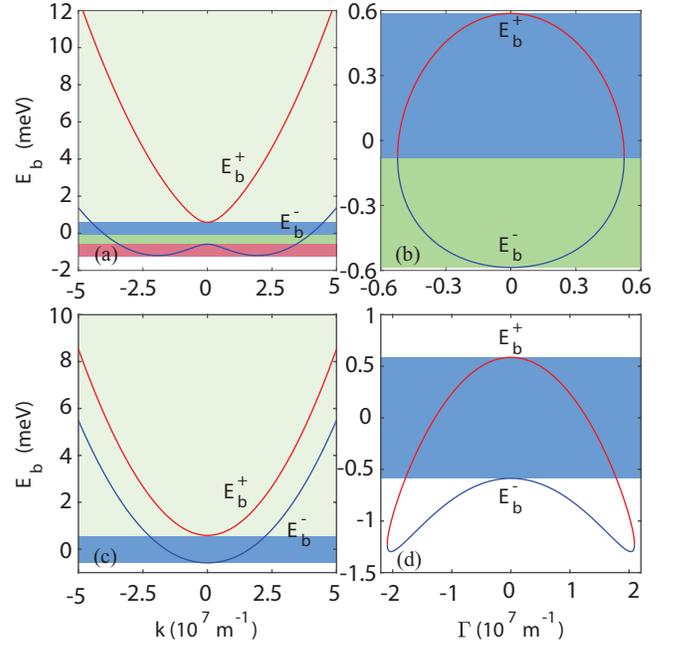}
\caption{\label{fig_bulkspectrum}The bulk spectrum of the KP superlattice model with SOC. (a) The bulk spectrum of the plane-wave solution (\ref{eq_bulkspectrum}) in the strong SOC regime $x_{\rm so}=50$ nm. (b) The bulk spectrum of the exponential function solution (\ref{eq_bulkspectrumII}) in the strong SOC regime $x_{\rm so}=50$ nm. (c) The bulk spectrum of the plane-wave solution (\ref{eq_bulkspectrum}) in the weak SOC regime $x_{\rm so}=200$ nm. (d) The bulk spectrum of the exponential function solution (\ref{eq_bulkspectrumII}) in the weak SOC regime $x_{\rm so}=200$ nm}
\end{figure}

Second, we consider the exponential function solution. The bulk wave function now can be assumed as
\begin{equation}
\Psi_{\rm b}(x)=e^{-\Gamma\,x}\left(
\begin{array}{c}
\chi_{1} \\
\chi_{2}
\end{array}
\right),
\end{equation}
where $\chi_{1,2}$ are also the coefficients to be determined. The bulk Schr\"odinger equation $(H_{\rm b}-E_{\rm b})\Psi_{\rm b}=0$ now gives us
\begin{equation}
\left(\begin{array}{cc}\frac{-\Gamma^{2}}{2m}-E_{\rm b}+i\alpha\Gamma&\Delta\\\Delta&\frac{-\Gamma^{2}}{2m}-E_{\rm b}-i\alpha\Gamma\end{array}\right)\cdot\left(\begin{array}{c}\chi_{1}\\\chi_{2}\end{array}\right)=0.\label{eq_bulkschrodinger}
\end{equation}
Setting the determinant of the matrix (on the left side of the above equation) equal to zero, we get the bulk energy
\begin{equation}
E^{\pm}_{\rm b}=-\frac{\Gamma^{2}}{2m}\pm\sqrt{-\alpha^{2}\Gamma^{2}+\Delta^{2}}.\label{eq_bulkspectrumII}
\end{equation}
Substituting the bulk energy in the bulk Schr\"odinger equation (\ref{eq_bulkschrodinger}) with the above results, we obtain the corresponding bulk wave functions:
\begin{equation}
\Psi^{+}_{\rm b}=\left\{\begin{array}{c}e^{-\Gamma\,x}\left(\begin{array}{c}e^{i\varphi}\\1\end{array}\right)\\
e^{\Gamma\,x}\left(\begin{array}{c}e^{-i\varphi}\\1\end{array}\right)\end{array}\right.,
\Psi^{-}_{\rm b}=\left\{\begin{array}{c}e^{-\Gamma\,x}\left(\begin{array}{c}-e^{-i\varphi}\\1\end{array}\right)\\
e^{\Gamma\,x}\left(\begin{array}{c}-e^{i\varphi}\\1\end{array}\right)\end{array}\right.,\label{eq_branchwave2}
\end{equation}
where $\varphi\equiv\varphi(\Gamma)=\arctan\left(\alpha\Gamma/\sqrt{-\alpha^{2}\Gamma^{2}+\Delta^{2}}\right)$. Because $E^{\pm}_{\rm b}$ is still an even function of $\Gamma$, there are two degenerate bulk wave functions, i.e., the $\Gamma$ solution and the $-\Gamma$ solution.

All of the above results are very general. Here, taking the InSb nanowire material as a concrete example, we show the bulk spectrum of the spin-orbital superlattice model in Fig.~\ref{fig_bulkspectrum}. Figures~\ref{fig_bulkspectrum}(a) and (b) show the bulk spectrum in the strong SOC regime ($m\alpha^{2}>g_{e}\mu_{B}B$), and Figs.~\ref{fig_bulkspectrum}(c) and (d) show the bulk spectrum in the weak SOC regime ($m\alpha^{2}<g_{e}\mu_{B}B$). Also, from the detailed expressions of the bulk spectrum (\ref{eq_bulkspectrum}) and (\ref{eq_bulkspectrumII}), we can derive some general results that are very useful in the following calculations. In the strong SOC regime [see Figs.~\ref{fig_bulkspectrum}(a) and (b)], $E^{+}_{\rm b}>\Delta$ and $E^{-}_{\rm b}>-\frac{1}{2}m\alpha^{2}-\frac{\Delta^{2}}{2m\alpha^{2}}$ for the plane-wave solution, and $-\Delta\le\,E^{-}_{\rm b}\le\,-\frac{\Delta^{2}}{2m\alpha^{2}}$ and $-\frac{\Delta^{2}}{2m\alpha^{2}}\le\,E^{+}_{\rm b}\le\Delta$ for the exponential function solution. In the weak SOC regime [see Figs.~\ref{fig_bulkspectrum}(c) and (d)], $E^{+}_{\rm b}>\Delta$ and $E^{-}_{\rm b}>-\Delta$ for the plane-wave solution, and $-\frac{1}{2}m\alpha^{2}-\frac{\Delta^{2}}{2m\alpha^{2}}\le\,E^{-}_{\rm b}\le\,-\Delta$ and $-\frac{\Delta^{2}}{2m\alpha^{2}}\le\,E^{+}_{\rm b}\le\Delta$ for the exponential function solution.

\section{The transcendental equation}\label{appendix_B}
Each bulk wave function can not satisfy the desired boundary conditions [see Eqs.~(\ref{eq_boundary_I}) and (\ref{eq_boundary_II})]. However, a linear combination of the four degenerate bulk wave functions can fulfill the boundary conditions~\cite{Bulgakov,Tsitsishvili}. Here, in different energy regions when the system is in the strong SOC regime ($m\alpha^{2}>g_{e}\mu_{B}B$), we totally derive four transcendental equations,  i.e., Eqs.~(\ref{eq_transc1}), (\ref{eq_transc2}), (\ref{eq_transc3}), and (\ref{eq_transc4}), which represent the implicit relation between the energy and the Bloch wave vector. It should be noted that two of the equations, i.e., Eqs.~(\ref{eq_transc2}) and (\ref{eq_transc4}), also are valid in the weak SOC regime ($m\alpha^{2}<g_{e}\mu_{B}B$).

\subsection{The region: $-\Delta>E>-\frac{1}{2}m\alpha^{2}-\frac{\Delta^{2}}{2m\alpha^{2}}$}
As can be seen from the bulk spectrum of the Hamiltonian (\ref{eq_Hamiltonain}) [see Fig.~\ref{fig_bulkspectrum}(a)], when $-\Delta>E>-\frac{1}{2}m\alpha^{2}-\frac{\Delta^{2}}{2m\alpha^{2}}$, one can solve four $k$ solutions, i.e., $\pm\,k_{1,2}$, from the `$-$' branch dispersion relation given in Eq.~(\ref{eq_bulkspectrum})
\begin{equation}
k_{1,2}=\sqrt{2}m\alpha\sqrt{1+\frac{E}{m\alpha^{2}}\pm\sqrt{1+2\frac{E}{m\alpha^{2}}+\frac{\Delta^{2}}{m^{2}\alpha^{4}}}}.
\end{equation}
The eigenfunction $\Psi(x)$ of Hamiltonian (\ref{eq_Hamiltonain}) can be expanded in terms of these four bulk wave functions. All we need to do is to let the expanded eigenfunction $\Psi(x)$ satisfy the desired boundary conditions [see Eqs.~(\ref{eq_boundary_I}) and (\ref{eq_boundary_II})]. It should be noted that all of the four bulk wave functions belong to the `$-$' branch $\Psi^{-}_{\rm b}(x)$ given in Eq.~(\ref{eq_branchwave1}). Thus, in the coordinate region $0<x<a$, the eigenfunction $\Psi(x)$ can be written as follows
\begin{eqnarray}
\Psi(x)&=&c_{1}\left(
\begin{array}{c}
\sin\frac{\theta_{1}}{2}   \\
-\cos\frac{\theta_{1}}{2}
\end{array}
\right)e^{ik_{1}x}+c_{2}\left(
\begin{array}{c}
\cos\frac{\theta_{1}}{2}   \\
-\sin\frac{\theta_{1}}{2}
\end{array}
\right)e^{-ik_{1}x}+\nonumber\\
&&c_{3}\left(
\begin{array}{c}
\sin\frac{\theta_{2}}{2}   \\
-\cos\frac{\theta_{2}}{2}
\end{array}
\right)e^{ik_{2}x}+c_{4}\left(
\begin{array}{c}
\cos\frac{\theta_{2}}{2}   \\
-\sin\frac{\theta_{2}}{2}
\end{array}
\right)e^{-ik_{2}x},\nonumber\\\label{eq_generalwfun1}
\end{eqnarray}
where $\theta_{1,2}\equiv\theta_{1,2}(k)=\arctan[\Delta/(\alpha\,k_{1,2})]$, and $c_{1,2,3,4}$ are the coefficients to be determined. In the region $-a<x<0$, the eigenfunction can be written down with the help of the Bloch theorem,
\begin{equation}
\Psi(x)=e^{-ik_{b}a}\Psi(x+a),
\end{equation}
where we have introduced the Bloch wave vector $k_{b}=l\pi/(Na)$ ($l=-N/2,\cdots,N/2$). In the above equation, because $0<x+a<a$, the right side of the above equation can be expressed with the help of Eq.~(\ref{eq_generalwfun1}). Now, we consider the boundary conditions at the site $x=0$. Substituting the wavefunction $\Psi(x)$ and the first derivative $\Psi'(x)$ in Eqs.~(\ref{eq_boundary_I}) and (\ref{eq_boundary_II}) with the above-derived expressions, we obtain an equation array,
\begin{equation}
{\bf M}_{1}\cdot{\bf C}=0,
\end{equation}
where ${\bf M}_{1}$ is a $4\times4$ matrix, and ${\bf C}=(c_{1},c_{2},c_{3},c_{4})^{\rm T}$. Letting the determinant of the matrix ${\bf M}_{1}$ equal to $0$, i.e., ${\rm det}({\bf M}_{1})=0$, we have the following transcendental equation
\begin{widetext}
\begin{eqnarray}
&&\big[2k_{1}k_{2}(1-\sin\theta_{1}\sin\theta_{2})-(k^{2}_{1}+k^{2}_{2})\cos\theta_{1}\cos\theta_{2}\big]\sin[(k_{1}+k_{b})a/2]\sin[(k_{1}-k_{b})a/2]\sin[(k_{2}+k_{b})a/2]\sin[(k_{2}-k_{b})a/2]\nonumber\\
&&~~~~~~~~~~~~~~-k_{0}\big(k_{1}-k_{1}\sin\theta_{1}\sin\theta_{2}-k_{2}\cos\theta_{1}\cos\theta_{2}\big)\sin[k_{2}a]\sin[(k_{1}+k_{b})a/2]\sin[(k_{1}-k_{b})a/2]\nonumber\\
&&~~~~~~~~~~~~~~-k_{0}\big(k_{2}-k_{2}\sin\theta_{1}\sin\theta_{2}-k_{1}\cos\theta_{1}\cos\theta_{2}\big)\sin[k_{1}a]\sin[(k_{2}+k_{b})a/2]\sin[(k_{2}-k_{b})a/2]\nonumber\\
&&~~~~~~~~~~~~~~+k^{2}_{0}\left(\sin^{2}[(k_{1}+k_{2})a/2]\sin^{2}[(\theta_{1}-\theta_{2})/2]-\sin^{2}[(k_{1}-k_{2})a/2]\cos^{2}[(\theta_{1}+\theta_{2})/2]\right)=0,\label{eq_transc1}
\end{eqnarray}
\end{widetext}
where $k_{0}=mV_{0}a$. This equation actually is an implicit relation between the energy $E$ and the Bloch wavevector $k_{b}$.

\subsection{The region: $E>\Delta$}
As can also be seen from the bulk spectrum of Hamiltonian (\ref{eq_Hamiltonain}) [see Fig.~\ref{fig_bulkspectrum}(a)], when $E>\Delta$, one can solve two solutions $\pm\,k_{1}$ from the `$+$' branch and
two solutions $\pm\,k_{2}$ from the `$-$' branch dispersion relations given in Eq.~(\ref{eq_bulkspectrum}).
Therefore, the eigenfunction $\Psi(x)$ can be expanded in terms of the four bulk wave functions, i.e., two from the `$+$' branch and two from the `$-$' branch given in Eq.~(\ref{eq_branchwave1}). In the coordinate region $0<x<a$, the eigenfunction can be written as
\begin{eqnarray}
\Psi(x)&=&c_{1}\left(
\begin{array}{c}
\cos\frac{\theta_{1}}{2}   \\
\sin\frac{\theta_{1}}{2}
\end{array}
\right)e^{ik_{1}x}+c_{2}\left(
\begin{array}{c}
\sin\frac{\theta_{1}}{2}   \\
\cos\frac{\theta_{1}}{2}
\end{array}
\right)e^{-ik_{1}x}+\nonumber\\
&&c_{3}\left(
\begin{array}{c}
\sin\frac{\theta_{2}}{2}   \\
-\cos\frac{\theta_{2}}{2}
\end{array}
\right)e^{ik_{2}x}+c_{4}\left(
\begin{array}{c}
\cos\frac{\theta_{2}}{2}   \\
-\sin\frac{\theta_{2}}{2}
\end{array}
\right)e^{-ik_{2}x}.\nonumber\\
\end{eqnarray} 
Repeating the same procedures as those given in the above subsection, we obtain the following transcendental equation
\begin{widetext}
\begin{eqnarray}
&&\left(2k_{1}k_{2}(1+\sin\theta_{1}\sin\theta_{2})+(k^{2}_{1}+k^{2}_{2})\cos\theta_{1}\cos\theta_{2}\right)\sin[(k_{1}+k_{b})a/2]\sin[(k_{1}-k_{b})a/2]\sin[(k_{2}+k_{b})a/2]\sin[(k_{2}-k_{b})a/2]\nonumber\\
&&~~~~~~~~~~~~~~-k_{0}\big(k_{1}+k_{1}\sin\theta_{1}\sin\theta_{2}+k_{2}\cos\theta_{1}\cos\theta_{2}\big)\sin[k_{2}a]\sin[(k_{1}+k_{b})a/2]\sin[(k_{1}-k_{b})a/2]\nonumber\\
&&~~~~~~~~~~~~~~-k_{0}\big(k_{2}+k_{2}\sin\theta_{1}\sin\theta_{2}+k_{1}\cos\theta_{1}\cos\theta_{2}\big)\sin[k_{1}a]\sin[(k_{2}+k_{b})a/2]\sin[(k_{2}-k_{b})a/2]\nonumber\\
&&~~~~~~~~~~~~~~+k^{2}_{0}\big(\sin^{2}[(k_{1}+k_{2})a/2]\cos^{2}[(\theta_{1}-\theta_{2})/2]-\sin^{2}[(k_{1}-k_{2})a/2]\sin^{2}[(\theta_{1}+\theta_{2})/2]\big)=0.\label{eq_transc2}
\end{eqnarray}
\end{widetext}
This equation actually is an implicit relation between the energy $E$ and the Bloch wavevector $k_{b}$.

\subsection{The region: $-\Delta<E<-\frac{\Delta^{2}}{2m\alpha^{2}}$}
So far, only the bulk plane-wave solutions are used for solving the energy band of our model (\ref{eq_Hamiltonain}). Here, we move to consider another interesting energy region, where $E$ lies inside the bulk energy gap. In this case, we have to utilize the exponential function solutions.

In the energy region $-\Delta<E<-\frac{\Delta^{2}}{2m\alpha^{2}}$, one can find two solutions $\pm\,k$ from the `$-$' branch dispersion relation given in Eq.~(\ref{eq_bulkspectrum})
\begin{equation}
k=\sqrt{2}m\alpha\sqrt{1+\frac{E}{m\alpha^{2}}+\sqrt{1+2\frac{E}{m\alpha^{2}}+\frac{\Delta^{2}}{m^{2}\alpha^{4}}}}.
\end{equation}
One can also find two solutions $\pm\Gamma$ from the `$-$' branch dispersion relation given in Eq.~(\ref{eq_bulkspectrumII})
\begin{equation}
\Gamma=\sqrt{2}m\alpha\sqrt{-1-\frac{E}{m\alpha^{2}}+\sqrt{1+2\frac{E}{m\alpha^{2}}+\frac{\Delta^{2}}{m^{2}\alpha^{4}}}}.
\end{equation}
Thus, the eigenfunction $\Psi(x)$ of Hamiltonian (\ref{eq_Hamiltonain}) can be expanded in terms of these four bulk wave functions, i.e., two from the `$-$' branch of the plane-wave solution and two from the `$-$' branch of the exponential function solution. In the coordinate region $0<x<a$, the eigenfunction can be expanded as
\begin{eqnarray}
\Psi(x)&=&c_{1}e^{-\Gamma\,x}\left(\begin{array}{c}-e^{-i\varphi}\\1\end{array}\right)+c_{2}e^{\Gamma\,x}\left(\begin{array}{c}-e^{i\varphi}\\1\end{array}\right)+\nonumber\\
&&c_{3}e^{ikx}\left(\begin{array}{c}\sin\frac{\theta}{2}\\-\cos\frac{\theta}{2}\end{array}\right)+c_{4}e^{-ikx}\left(\begin{array}{c}\cos\frac{\theta}{2}\\-\sin\frac{\theta}{2}\end{array}\right),
\end{eqnarray}
where $\theta\equiv\theta(k)=\arctan\left[\Delta/(\alpha\,k)\right]$, $\varphi\equiv\varphi(\Gamma)=\arctan\left(\alpha\Gamma/\sqrt{-\alpha^{2}\Gamma^{2}+\Delta^{2}}\right)$, and $c_{1,2,3,4}$ are the expansion coefficients to be determined. Repeating the same procedures as those given in the first subsection,
we obtain the following transcendental equation
\begin{widetext}
\begin{eqnarray}
&&\big[(k^{2}-\Gamma^{2})\cos\theta\sin\varphi-2k\Gamma(\cos\varphi-\sin\theta)\big]\big(\cos[k_{b}a]-\cosh[\Gamma\,a]\big)\sin[(k+k_{b})a/2]\sin[(k-k_{b})a/2]\nonumber\\
&&~~~~~~~~~~~~~~~~~~+k_{0}\big(\Gamma\cos\varphi-\Gamma\sin\theta-k\cos\theta\sin\varphi\big)\big(\cos[k_{b}a]-\cosh[\Gamma\,a]\big)\sin[ka]\nonumber\\
&&~~~~~~~~~~~+2k_{0}\big(k\cos\varphi-k\sin\theta+\Gamma\cos\theta\sin\varphi\big)\sinh[\Gamma\,a]\sin[(k+k_{b})a/2]\sin[(k-k_{b})a/2]\nonumber\\
&&~~~~~~~~~~~~-k^{2}_{0}\cos\theta\sin\varphi\big(\cos[ka]\cosh[\Gamma\,a]-1\big)-k^{2}_{0}\big(\cos\varphi-\sin\theta\big)\sinh[\Gamma\,a]\sin[ka]=0.\label{eq_transc3}
\end{eqnarray}
\end{widetext}
This equation actually is an implicit relation between the energy $E$ and the Bloch wavevector $k_{b}$.

\subsection{The region: $-\frac{\Delta^{2}}{2m\alpha^{2}}<E<\Delta$}
In this energy region, one can find two solutions $\pm\,k$ from the `$-$' branch dispersion relation given in Eq.~(\ref{eq_bulkspectrum})
and two solutions $\pm\Gamma$ from the `$+$' branch dispersion relation given in Eq.~(\ref{eq_bulkspectrumII}).
Thus, the eigenfunction $\Psi(x)$ of Hamiltonian (\ref{eq_Hamiltonain}) can be expanded in terms of these four bulk wavefunctions, i.e., two from the `$-$' branch of the plane-wave solution and two from the `$+$' branch of the exponential function solution. Now, in the coordinate region $0<x<a$, the eigenfunction can be expanded generally as
\begin{eqnarray}
\Psi(x)&=&c_{1}e^{-\Gamma\,x}\left(\begin{array}{c}e^{i\varphi}\\1\end{array}\right)+c_{2}e^{\Gamma\,x}\left(\begin{array}{c}e^{-i\varphi}\\1\end{array}\right)+\nonumber\\
&&c_{3}e^{ikx}\left(\begin{array}{c}\sin\frac{\theta}{2}\\-\cos\frac{\theta}{2}\end{array}\right)+c_{4}e^{-ikx}\left(\begin{array}{c}\cos\frac{\theta}{2}\\-\sin\frac{\theta}{2}\end{array}\right).
\end{eqnarray}
Repeating the same procedures as those given in the first subsection, we obtain the following transcendental equation
\begin{widetext}
\begin{eqnarray}
&&\big[(k^{2}-\Gamma^{2})\cos\theta\sin\varphi+2k\Gamma(\cos\varphi+\sin\theta)\big]\big(\cos[k_{b}a]-\cosh[\Gamma\,a]\big)\sin[(k+k_{b})a/2]\sin[(k-k_{b})a/2]\nonumber\\
&&~~~~~~~~~~~~~~~~~~-k_{0}\big(\Gamma(\cos\varphi+\sin\theta)+k\cos\theta\sin\varphi\big)\big(\cos[k_{b}a]-\cosh[\Gamma\,a]\big)\sin[ka]\nonumber\\
&&~~~~~~~~~~~-2k_{0}\big(k(\cos\varphi+\sin\theta)-\Gamma\cos\theta\sin\varphi\big)\sinh[\Gamma\,a]\sin[(k+k_{b})a/2]\sin[(k-k_{b})a/2]\nonumber\\
&&~~~~~~~~~~~~-k^{2}_{0}\cos\theta\sin\varphi\big(\cos[ka]\cosh[\Gamma\,a]-1\big)+k^{2}_{0}\big(\cos\varphi+\sin\theta\big)\sinh[\Gamma\,a]\sin[ka]=0.\label{eq_transc4}
\end{eqnarray}
\end{widetext}
This equation actually is an implicit relation between the energy $E$ and the Bloch wavevector $k_{b}$.

We now consider the transcendental equations when the superlattice system is in the weak SOC regime [see Figs.~\ref{fig_bulkspectrum}(c) and (d)]. In the weak SOC regime, the energy region is only divided into two parts, i.e., $E>\Delta$ and $\Delta>E>-\Delta$ [see Figs.~\ref{fig_bulkspectrum}(c) and (d)]. The discussions in this regime are just the same as those in the strong SOC regime. In the energy region $E>\Delta$, the transcendental equation still has the form given in Eq.~(\ref{eq_transc2}). In the energy region $\Delta>E>-\Delta$, the transcendental equation is also exactly the same as that given in Eq.~(\ref{eq_transc4}).

\end{document}